\documentclass{article}

\usepackage{microtype}
\usepackage{graphicx}
\usepackage{subfigure}
\usepackage{booktabs} 
\usepackage{xspace} 
\usepackage{adjustbox}

\usepackage{subcaption}
\usepackage{multirow}
\usepackage{float} 
\usepackage{caption}
\usepackage{subcaption} 
\usepackage{amsmath}

\usepackage{hyperref}

\newcommand{\name}{\textit{DeepSync}\xspace} 

\usepackage[accepted]{icml2025}

\usepackage{amsmath}
\usepackage{amssymb}
\usepackage{mathtools}
\usepackage{amsthm}

\usepackage{textcomp}

\usepackage[capitalize,noabbrev]{cleveref}

\usepackage{comment}
\newcommand{\squishlist}{
	\begin{list}{$\bullet$}
		{  \setlength{\leftmargin}{+0.15in}
		} }

		\newcommand{\squishend}{
	\end{list} }
    
\theoremstyle{plain}

\theoremstyle{definition}

\theoremstyle{remark}

\usepackage[textsize=tiny]{todonotes}

\begin{document}

\twocolumn[
\icmltitle{\name: A Learning Framework for Pervasive Localization using Code Synchronization on Compressed Cellular Spectrum}



\icmlsetsymbol{equal}{*}

\begin{icmlauthorlist}
\icmlauthor{Aritrik Ghosh}{a}
\icmlauthor{Nakul Garg}{a}
\icmlauthor{Nirupam Roy}{a}
\end{icmlauthorlist}

\icmlaffiliation{a}{ 1. Department of Computer Science, University of Maryland College Park}



\icmlkeywords{Machine Learning, ICML}

\vskip 0.3in
]



\printAffiliationsAndNotice{}  

\begin{abstract}
Pervasive localization is essential for continuous tracking applications, yet existing solutions face challenges in balancing power consumption and accuracy. GPS, while precise, is impractical for continuous tracking of micro-assets due to high power requirements. Recent advances in non-linear compressed spectrum sensing offer low-power alternatives, but existing implementations achieve only coarse positioning through Received Signal Strength Indicator (RSSI) measurements. We present \name, a deep learning framework that enables precise localization using compressed cellular spectrum. Our key technical insight lies in formulating sub-sample timing estimation as a template matching problem, solved through a novel architecture combining temporal CNN encoders for multi-frame processing with cross-attention mechanisms. The system processes non-linear inter-modulated spectrum through hierarchical feature extraction, achieving robust performance at SNR levels below -10dB - a regime where conventional timing estimation fails. By integrating real cellular infrastructure data with physics-based ray-tracing simulations, \name achieves 2.128-meter median accuracy while consuming significantly less power than conventional systems. Real-world evaluations demonstrate $10\times$ improvement over existing compressed spectrum approaches, establishing a new paradigm for ultra-low-power localization.
\vspace{-0.2in}
\end{abstract}

\section{Introduction}

$\blacksquare$ {\bf Motivation.}
Pervasive localization underpins numerous applications requiring continuous position tracking - from monitoring elderly patients and pets to securing personal belongings. While GPS has established itself as the dominant positioning technology, its significant power consumption makes it impractical for energy-constrained scenarios demanding persistent tracking. Cellular networks, with their dense constellation of geographically distributed base stations, present a compelling opportunity for low-power localization by leveraging existing infrastructure.

$\blacksquare$ {\bf Challenges.}
Leveraging cellular signals for low-power localization faces three fundamental challenges. First, traditional decoding of cellular signals demands significant power consumption due to RF-to-baseband downconversion \cite{hentschel2002digital}. Second, the wide cellular spectrum, spanning multiple frequency bands, requires sequential scanning by typical low-cost receivers, introducing substantial latency \cite{dahlman20164g}. Recently, a breakthrough in low-power wireless reception has emerged through compressed spectrum sensing and non-linear signal processing \cite{li2015w, rostami2021mixiq, ensworth2017low, guo2022saiyan}, demonstrating capabilities to decode various wireless protocols including WiFi, Bluetooth, and cellular signals while consuming 100$\times$ less power than conventional receivers. While these emerging architectures have addressed the power and latency constraints through passive mixing and intermodulation, existing implementations achieve only coarse positioning through Received Signal Strength Indicator (RSSI) measurements \cite{garg2024litefoot}, resulting in sub-optimal localization accuracy. These limitations have prevented cellular-based systems from achieving the GPS-like precision necessary for practical asset tracking applications.

Achieving precise localization fundamentally relies on timing information extracted from synchronization codes - specialized signal templates embedded within cellular transmissions \cite{3gpp_tr_22_872}. At typical cellular sampling rates, even a single sample timing error can result in localization errors exceeding 150 meters due to the speed of light, necessitating sub-sample timing precision. 
However, in compressed spectrum architectures, these synchronization codes undergo severe degradation through non-linear intermodulation and multi-band interference \cite{rostami2021mixiq, garg2024litefoot}, where signal components across multiple frequency bands mix unpredictably. This results in unprecedented signal-to-noise ratio (SNR) challenges, with received synchronization signals experiencing 20-30 dB lower SNR compared to traditional cellular systems - equivalent to 100-1000$\times$ worse signal quality. Such extreme SNR degradation renders conventional timing estimation techniques ineffective.


$\blacksquare$ {\bf Our Approach.}
We present \name, a deep learning system that enables precise localization using compressed cellular spectrum. Our key technical insight lies in formulating sub-sample timing estimation as a template matching problem, solved through a novel architecture with three key components: a cross-attention mechanism for precise template alignment, a temporal CNN encoder leveraging multiple consecutive frames, and the target sync-code encoder for robust sync-code detection. This architecture enables accurate timing offset estimation from non-linear compressed signals, facilitating precise time-difference-of-arrival measurements. Drawing from template matching advances in computer vision and long-term video object tracking \cite{Wu_2024_CVPR, ye2022joint, Mayer_2022_CVPR}, our system achieves robust timing estimation in previously intractable SNR levels. \name achieves 2.182-meter median accuracy while consuming $50\times$ less power consumption compared to traditional positioning systems—without requiring any infrastructure modifications. This represents an advancement over the compressed spectrum approaches that achieve 20-meter accuracy using RSSI measurements.

$\blacksquare$ {\bf Contributions.}
Our primary contribution is a neural architecture for precise sub-sample timing estimation in intermodulated cellular spectra. \name combines a dual CNN-based encoder with cross-attention mechanisms, processing non-linear spectral components through hierarchical convolutional layers for multi-scale feature extraction. This architecture estimates sub-sample offsets from highly compressed signals by leveraging the preserved orthogonal properties of sync-codes after spectrum folding. To train the model, we developed an RF ray-tracing based data generation framework that integrates open-source cellular infrastructure data (cell tower coordinates, identifiers) with urban geometry and material properties. Our training methodology exploits the periodicity of cellular frames through temporal aggregation, enabling accurate offset estimation at Signal-to-noise ratio (SNR) below -10dB. The key contributions of this work can be summarized as:
\vspace{-0.05in}

    
    
    
\squishlist
    \item A cross-attention based architecture for time-difference-of-arrival estimation in intermodulated spectra, achieving 2.128-meter median localization accuracy.
    \vspace{-0.05in}
    
    \item A physics-informed digital-twin framework integrating real cellular infrastructure data with electromagnetic propagation models for synthetic data generation.
    \vspace{-0.05in}
    
    \item A novel temporal aggregation method for multi-frame processing, enhancing performance in compressed non-linear spectra.
    \vspace{-0.05in}
    
    \item Real-world evaluations demonstrating $10\times$ accuracy improvement over existing cellular localization systems while maintaining ultra-low power consumption.
\squishend

\section{Preliminaries}
\vspace{-0.1in}
In this section we will provide a primer on cellular localization and sub-sample template matching.

$\blacksquare$ \textbf{Basics of Localization.}
Localization determines an object's position by measuring distances from multiple fixed reference points, called anchors. These distances are typically estimated through three popular approaches: Received Signal Strength Indicator (RSSI), Time of Arrival (TOA), or Time Difference of Arrival (TDOA). RSSI estimates distance by correlating signal strength attenuation with distance from the anchor point, often modeled using path-loss equations. In TOA, distance $d$ is calculated from signal travel time $t$ as $d = ct$, where $c$ is the speed of light. TDOA uses differences in arrival times between anchor pairs to compute distance differences $\Delta d_{ij} = d_i - d_j = c(t_i - t_j)$, forming hyperbolic curves whose intersections indicate the object's location. Once distances to at least three anchors in 2D are known, the object's position can be estimated through trilateration.

$\blacksquare$ \textbf{Trilateration.}
Consider N anchor points with known positions $(x_i,y_i)$, $i \in N$ in 2D space. When signal transmission time is known, the distance $d_i$ between an anchor and unknown point $(x,y)$ is computed using time of arrival as $d = c(t_{arrival}-t_{sent})$, where $c$ is the speed of light. These distances form circles around each anchor described by $d_i = \sqrt{(x_i-x)^2+(y_i-y)^2}$, whose intersection reveals the target position. However, when transmission time is unknown, we leverage time differences between anchor pairs. The time difference of arrival (TDOA) between two anchors $i$ and $j$ yields $\Delta t_{ij} = \frac{d_i - d_j}{c}$, corresponding to distance difference $\Delta d_{ij} = d_i - d_j = c\Delta t_{ij}$. These differences create hyperbolic equations $\sqrt{(x - x_i)^2 + (y - y_i)^2} - \sqrt{(x - x_j)^2 + (y - y_j)^2} = \Delta d_{ij}$, $i, j \in N$, $i \neq j$, whose intersections determine the target position. The accuracy of this position estimation critically depends on precise measurement of time differences $\Delta t_{ij}$. In practical systems, timing is measured through sample offsets $\tau$, where each offset corresponds to a distance $d = \frac{\tau c}{f}$, with $f$ being the sampling frequency.

$\blacksquare$ \textbf{Importance of code synchronization for localization.}
Precise timing estimation, critical for TDOA-based localization, relies on detecting sync-codes embedded within cellular transmissions. These sync-codes are carefully designed sequences with known patterns that enable receivers to align with transmitted signals and extract accurate timing information. In cellular networks, Primary Synchronization Signal (PSS) and Secondary Synchronization Signal (SSS) serve as sync-codes for this purpose. At typical cellular sampling rates of 1.92 MHz, even a single sample timing error translates to a position error of 156.25m. Sub-sample precision is therefore crucial - ideally, timing errors should be within 0.1 samples to achieve meter-level accuracy. The fundamental challenge lies in extracting precise timing information from these sync-codes in practical deployments where signals experience various forms of degradation and interference.

$\blacksquare$ \textbf{Signal Processing in Compressed Spectrum.}
Wireless signals, including cellular transmissions, are carried at high frequencies through multiple subcarriers. Traditional receivers use complex circuitry to downconvert these high-frequency signals to lower frequencies for processing. An alternative approach leverages non-linear transformations, where signals at different frequencies naturally multiply with each other. When two signals at frequencies $f_1$ and $f_2$ undergo non-linear processing (like squaring), they produce components at their sum $(f_1 + f_2)$ and difference $(f_1 - f_2)$ frequencies. This property enables automatic downconversion of high-frequency signals to baseband frequencies near zero, a phenomenon known as spectrum folding. In cellular systems using multiple subcarriers, this non-linear mixing creates intermodulation - where all subcarrier frequencies interact, producing a compressed version of the original spectrum at baseband. While this approach dramatically simplifies receiver design, it introduces significant challenges for sync-code detection. The sync-codes become embedded within this compressed, intermodulated spectrum where they experience severe SNR degradation due to interference from other frequency components. Extracting precise timing information from these degraded sync-codes becomes particularly challenging, as the non-linear transformation fundamentally alters the signal structure while mixing noise across the spectrum. This challenge forms the core technical problem addressed in this work.

\section{Challenges and Intuitions}

$\blacksquare$ \textbf{Noisy Signal Spectrum.}
\name's receiver architecture employs non-linear signal processing (specifically squaring) for power-efficient downconversion. While this enables simultaneous wide-band mixing through spectrum folding, it introduces significant challenges in detecting sync-codes (PSS and SSS) at baseband. The folded spectrum contains intermodulated components from the entire LTE band, where sync-codes become embedded within unpredictable mixing products from data subcarriers \cite{garg2024litefoot}. This degradation severely impacts sync-code detection - for a typical 15.36 MHz LTE signal, the recovered sync-code (bandwidth 1.4 MHz) experiences an SNR degradation of 20-30 dB compared to traditional systems. The problem compounds in practical deployments where multiple LTE bands with varying bandwidths are received simultaneously, resulting in 100-1000$\times$ worse signal quality through cross-band interference \cite{garg2024litefoot, rostami2021mixiq}.

$\blacksquare$ \textbf{Achieving Sub-sample offset in Noise.}
At typical cellular sampling rates, even a single sample timing error can result in localization errors exceeding 150 meters due to the speed of light \cite{3gpp_tr_22_872}. This challenge is exacerbated in low-SNR conditions where the correlation peak indicating sync-code alignment becomes broadened and less distinct due to factors like clock drift and inter-subcarrier mixing. Traditional correlation-based approaches are fundamentally limited to sample-level resolution \cite{nandakumar2016fingerio,vasisht2016decimeter}, making sub-sample precision crucial for accurate positioning. In TDOA-based systems, inconsistent sub-sample offsets across multiple base stations create incorrect hyperbolic intersections, causing substantial deviations in the estimated position. Such extreme SNR degradation through non-linear intermodulation and multi-band interference renders conventional timing estimation techniques ineffective, making reliable localization infeasible without precise sub-sample timing estimation in real-world deployments.

\section{Related Work}
\vspace{-0.1in}
\subsection{Learning in Communication and Localization}
\vspace{-0.1in}
Deep learning frameworks are leveraged to tackle various challenges in the wireless domain. Several studies have explored deep learning-based wireless channel estimation, demonstrating its potential in enhancing communication system reliability \cite{ge2021deep, karanam2023foundation, varshney2023deep, krijestorac2021spatial}. Concurrently, transformer-based architectures have been utilized for wireless protocol detection, as shown in \cite{belgiovine2024t}, while CNN-based models have been widely adopted for modulation classification and signal recognition \cite{schmidt2017wireless, jagannath2021multi}.

Beyond channel estimation and classification, neural networks have also been employed for signal decoding and de-mapping  \cite{he2019neural, schaedler2021recurrent}. In the context of target sync-code detection and synchronization, \cite{soltani2023pronto} introduced a CNN-based model to reduce Wi-Fi sync-code overhead while maintaining coarse frame synchronization. Additionally, studies such as \cite{ninkovic2020preamble, singh2024enhancements} explored code detection across different wireless standards, including Wi-Fi, LTE, and 5G.

Furthermore, recent advancements in source separation for OFDM systems have seen the adoption of deep learning techniques such as U-Net-based models \cite{lee2023neural} and diffusion models \cite{jayashankar2024score}. Neural Radiance Fields (NeRF)  \cite{lunewrf} demonstrating the use of NeRF to predict channel responses.

\subsection{Neural Networks in Template Matching}
\vspace{-0.1in}
Neural networks have traditionally been used in template matching, leveraging their ability to learn complex patterns. Convolutional Neural Networks (CNNs) have been used to enhance robustness against scale and rotation variations in template matching tasks \cite{bertinetto2016fully, danelljan2017eco, li2019siamrpn++}. Siamese networks, designed to learn similarity metrics, enable efficient comparisons between templates and target regions \cite{fan2019siamese, cheng2021learning, sun2020fast, shuai2021siammot}.

Transformer-based architectures have also been integrated into template matching frameworks. Two-stream two-stage trackers consist of two identical Transformer-based pipelines, separately extracting features from the target template and search region \cite{xie2021learning, lin2022swintrack}. Then One-stream One-stage trackers utilize a single Transformer-based pipeline, where feature extraction and fusion occur within the same network \cite{cui2022mixformer, chen2022backbone, ye2022joint}.

Additionally, hybrid models combine CNN-based feature extraction with Transformer-based feature matching, where CNN branches extract features from both the target template and search region, followed by a Transformer-driven similarity matching process \cite{chen2022hift, zhang2021fasttemplate, wang2021transformer}. 

\subsection{Exploiting Non-Linearity}
\vspace{-0.1in}
The use of non-linear signal properties has been a cornerstone of various communication systems. Non-linear backscatter circuits have enabled applications such as in-body localization \cite{vasisht2018body}. Similarly, harmonic RADARs and RFID-based systems have exploited non-linearity to mitigate environmental interference \cite{gomes2007use}.

\cite{garg2023sirius} adapted non-linearity for Angle of Arrival (AOA) estimation. Additionally, low-power radio receivers, such as those proposed in \cite{varshney2020tunnel, guo2022saiyan, ensworth2017low}, have utilized diode-based detectors to replace active components in receivers, improving energy efficiency. Envelope detectors for efficient downconversion have been demonstrated in \cite{rostami2021mixiq, 9858619, li2015w}. Additionally, emerging architectures leverage MEMS filters for signal conversion, which have been integrated into ultra-low-power backscatter communication systems \cite{jog2022enabling}.

\subsection{Cellular Localization}
\vspace{-0.1in}
Cellular localization can be achieved through various methods. RSSI-based localization in cellular systems provides an accuracy of 100–500 meters by leveraging signal strength and tower locations \cite{elbakly2019crescendo, ibrahim2011cellsense, sallouha2017localization}. This accuracy can be further improved to sub-50 meters using fingerprinting techniques \cite{heinrich2022airguard, laitinen2001database, sallouha2017localization}. Additionally, by combining RSSI with non-linearity, \cite{garg2024litefoot} achieved sub-20 meter accuracy in urban environments.

Time of Arrival (TOA) and Time Difference of Arrival (TDOA) methods offer further improvements, enabling sub-10 meter localization accuracy \cite{kim2017aoa, li2016toa, huang2019method, 9293258}. To enhance real-time and high-precision positioning, 5G-NR introduced a new reference signal called the Position Reference Signal (PRS) \cite{3gpp_tr_22_872}. Utilizing PRS, studies such as \cite{dwivedi2021positioning, muursepp2021performance, luo2023high, le20235g} have demonstrated sub-3 meter accuracy in localization.

\section{Problem Formulation}
\label{sec:problem_description}
\vspace{-0.1in}

In an ideal scenario, the PCI can be identified using matched filtering between the synchronization signal templates and the received signal. However, due to intermodulation and intramodulation interference, this task becomes non-trivial. The synchronization signal occupies only 6 resource blocks (RBs) in a typical 50 RB LTE frame. After spectrum folding, the signal SNR falls below \(-10\ \mathrm{dB}\), further degraded by intermodulation interference from other LTE bands.

To address this, we exploit the fact that synchronization signals repeat every \(10\ \mathrm{ms}\). By capturing \(N\) consecutive frames and stacking them, the random noise averages out while the synchronization signals reinforce. This process effectively improves the SNR by a factor of \(N\). For \(N = 20\), the improved SNR allows us to reliably extract the PCI using matched filtering on the aggregated signal.

\subsection{Coarse Synchronization: Sample Offset}
\vspace{-0.1in}
The sample offset indicates the coarse alignment of the synchronization signal within a frame. In the absence of interference, the offset can be determined using matched filtering. However, in practical scenarios with intermodulation and clock drift, matched filtering provides only a coarse estimate of the offset. Let the matched filter output be:
\[
\mathbf{r}[n] = \sum_{m=0}^{L-1} x[n+m] s[m],
\]
where \(x[n]\) is the received signal, \(s[m]\) is the synchronization signal template, and \(L\) is the template length. The index \(n^*\) corresponding to the peak of \(\mathbf{r}[n]\) gives the coarse offset. 

The coarse offset alone is insufficient for localization, as an error of even one sample (at \(1.92\ \mathrm{MHz}\) sampling rate) can result in a localization error of up to \(150\ \mathrm{m}\). Hence, sub-sample offset estimation is necessary.

\subsection{Fine Synchronization: Sub-sample Offset}
\vspace{-0.1in}
While matched filtering provides a coarse estimate, sub-sample offset estimation refines it to a higher precision. For a received frame sampled at \(1.92\ \mathrm{MHz}\), each frame has \(19200\) samples over \(10\ \mathrm{ms}\). Let the coarse estimate of a synchronization signal start at index \(i\). To ensure precision, we extract a segment of \(286\) samples centered around \(i\):
\[
\mathbf{x}[n] = \begin{cases} 
    x[n], & i-5 \leq n \leq i+280, \\
    0, & \text{otherwise}.
\end{cases}
\]

In this segment, multiple PCIs may overlap due to intermodulation. To disentangle their contributions, we employ a dual-network regression model trained to predict sub-sample offsets for each PCI. Given \(N\) detected PCIs, the model outputs \(N\) offset values \(\{\tau_k\}_{k=1}^N\).


\begin{figure*}[t]
    \centering
    \vspace{-0.1in}
    \includegraphics[width=6in, height=2.3in]{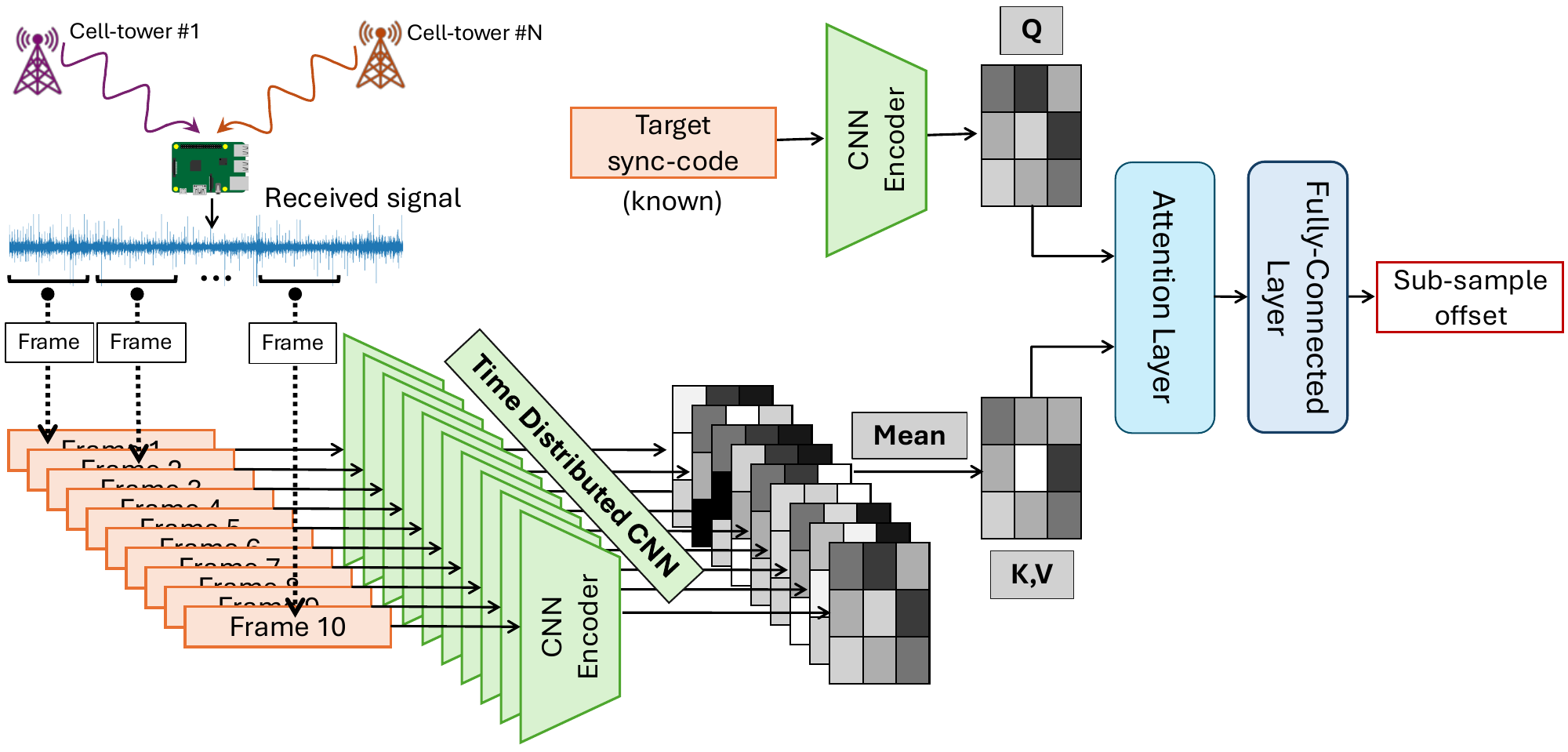}
    \vspace{-0.2in}
    \caption{\name's architecture}
    \vspace{-0.1in}
    \label{fig:model}
\end{figure*}

\vspace{-0.1in}
\section{Learning Architecture}
\label{sec:training}
\vspace{-0.1in}
Our proposed learning model is designed to predict the temporal offset of the target sync-code in the received compressed-spectrum signals. We aim to achieve sub-sample level accuracy in the predicted offset. This section elaborates on the learning architecture and signal pre-processing methods.

\subsection{Data Pre-processing}
\vspace{-0.1in}
In \name’s training pipeline, each sample consists of two primary inputs: a set of received signals and a target sync-code, both normalized to enhance numerical stability.

We capture 10 frames of cellular LTE signals and sum all frames together to improve the SNR of the sync-code \cite{garg2024litefoot}. Next, we perform coarse synchronization using standard correlation-based technique to identify the rough location of the target sync-code in the signal time series in term of a sample offset. Since coarse synchronization is not accurate, we extract this coarse sample offset along with 5 extra samples before and after the sync-code, resulting in a final length of 286 samples (for a 276-sample sync-code which is concatenated first PSS and SSS observed in a frame \cite{mathworks_lte_sync}). This estimated region is extracted from all 10 frames and passed through our model, along with the target sync-code, which was determined during coarse sample offset estimation.

To ensure stable optimization and prevent gradient explosion, we apply zero-mean, unit-variance normalization:

\begin{equation}
\mathbf{S}’ = \frac{\mathbf{S} - \mu_S}{\sigma_S}, \quad \mathbf{P}’ = \frac{\mathbf{P} - \mu_P}{\sigma_P},
\end{equation}

where $\mu_S, \sigma_S$ and $\mu_P, \sigma_P$ are computed over the training set. Wireless signals suffer from fading, noise, and hardware impairments \cite{stein1987fading,wu2017channel,zou2007compensation}, making raw signal processing unreliable. Normalization, widely adopted in wireless learning frameworks \cite{soltani2023pronto,zhou2019coarse,ninkovic2020preamble}, mitigates these variations and ensures consistent feature scaling.

\subsection{Model Description}
\vspace{-0.1in}
To accurately predict the temporal offset in received signals, \name integrates convolutional feature extraction and cross-attention-based alignment. The model comprises three main components: (1) a CNN-based signal encoder for extracting temporal features, (2) an independent encoder to represent the target sync-code signal, and (3) a cross-attention mechanism as shown in Figure \ref{fig:model}  to align signal representations and enhance feature matching.


This architecture is inspired by template matching paradigms widely used in computer vision, particularly in video tracking \cite{chen2021transformer,Mayer_2022_CVPR,cao2021hift,Wu_2024_CVPR} and object detection \cite{shahzad2021smart, zhou2021robust}. These methods take advantage of feature similarity computation to match a target template within a sequence of frames or within a image. Similarly, our approach learns to align received signals with the target sync-code.

$\blacksquare$ {\bf Time-distributed CNN encoders}
Our model consists of a dual CNN-based encoder designed to extract latent representations from both the received signals and the reference signal of the target sync-code. This choice is inspired by prior works \cite{soltani2023pronto, ninkovic2020preamble,he2021sounddet,zheng2024bat}, where convolutional architectures have demonstrated strong capabilities in capturing local temporal dependencies while maintaining computational efficiency.

The encoding function $f_\text{signal}(\cdot)$ is implemented as a deep stacked 1D CNN, where progressively smaller kernel sizes \((14, 7, 5)\) enable multi-scale feature extraction. Each convolutional layer \cite{o2015introduction} is followed by batch normalization \cite{ioffe2015batch} and ReLU activation \cite{agarap2018deep}, Finally, a Global Average Pooling (GAP) layer \cite{lin2013network} compresses the learned representations into a fixed-size embedding.

Each received signal $\mathbf{s}_i \in \mathbb{R}^{286 \times 1}$ is passed through this encoding function to extract a latent representation:
\begin{equation}
    \mathbf{z}_i = f_\text{signal}(\mathbf{s}_i), \quad \mathbf{z}_i \in \mathbb{R}^{158 \times 128}.
\end{equation}
To enhance robustness against SNR degradation and channel fading, feature maps across subsequent signals are aggregated using temporal averaging:
\begin{equation}
    \mathbf{Z} = \frac{1}{10} \sum_{i=1}^{10} \mathbf{z}_i, \quad \mathbf{Z} \in \mathbb{R}^{158 \times 128}.
\end{equation}

Similarly, the target sync-code $\mathbf{P}'$ is processed through an identical CNN encoder, ensuring feature representations from both input streams are spatially aligned:
\begin{equation}
    \mathbf{H} = f_\text{sync-code}(\mathbf{P}'), \quad \mathbf{H} \in \mathbb{R}^{158 \times 128}.
\end{equation}

$\blacksquare$ {\bf Cross-Attention Mechanism}
\vspace{-0.1in}

Attention mechanisms\cite{vaswani2017attention} have been widely adopted in wireless systems \cite{hamidi2021mcformer,guo2024parallel,belgiovine2024t}. We employ a cross-attention mechanism to align the received signal representations with the target sync-code, enhancing feature matching for offset estimation.

Our model leverages cross-attention to effectively align the received signal and target sync-code. Specifically, we treat the received signal as the query and the target sync-code as the key-value pair, enabling the network to dynamically learn temporal dependencies and selectively enhance features that contribute to accurate offset estimation.

The signal and target sync-code embeddings are combined using a multi-head cross-attention mechanism. Given the feature representations $\mathbf{Z} \in \mathbb{R}^{158 \times 128}$ from the signal encoder and $\mathbf{H} \in \mathbb{R}^{158 \times 128}$ from the target sync-code CNN encoder, attention scores are computed as:
\begin{equation}
    \text{Attention}(Q, K, V) = \text{softmax}\left(\frac{Q K^\top}{\sqrt{d_k}}\right) V.
\end{equation}
where $Q = Z W^Q, K = H W^K, V = H W^V$ and $W^Q, W^K, W^V $ are learned weight matrices, and $d_k$ is the key dimensionality.

The output of the attention block, $At$, represents the refined representation of the received signal after being aligned with the target sync-code. This is subsequently flattened and passed through fully connected layers to produce the final sample offset prediction:
\begin{equation}
    \hat{y} = f_\text{regression}(\text{Flatten}(At)),
\end{equation}
where $f_\text{regression}(\cdot)$ consists of 3 dense layers with ReLU activations and dropout.

$\blacksquare$ {\bf Loss Function and Optimization}
\vspace{-0.1in}

The model is trained using the Huber loss function to balance sensitivity to large errors while maintaining robustness to outliers:
\begin{equation}
    \mathcal{L}(y, \hat{y}) =
    \begin{cases} 
        \frac{1}{2}(y - \hat{y})^2, & \text{if } |y - \hat{y}| \leq \delta, \\
        \delta |y - \hat{y}| - \frac{1}{2} \delta^2, & \text{otherwise},
    \end{cases}
\end{equation}
where $\delta$ is a hyper-parameter set to $1.0$ in our experiments. 

The model is optimized using the Adam optimizer with a learning rate of $10^{-4}$, and the following metrics are monitored during training:
\begin{equation}
    \text{Mean Absolute Error (MAE): } \frac{1}{N} \sum_{i=1}^N |y_i - \hat{y}_i|.
\end{equation}

\vspace{-0.1in}
\section{Data Generation}
\vspace{-0.1in}
Building data from \cite{overpass-turbo} provides material properties, structural dimensions, and geographic coordinates essential for modeling signal propagation and attenuation. Cell tower locations and parameters obtained from \cite{cellmapper} are mapped to these building structures using geographic proximity algorithms. MATLAB's Communication Toolbox and ray-tracing models simulate LTE signal propagation through reconstructed environment.





Transmitter power is configured to 23 dBm for a 15.36 MHz bandwidth, following LTE standards \cite{ntia2010lte} for power and spectrum allocation.

Each transmitter-receiver pair is assigned a unique cell ID. This configuration enables multiple transmitters to serve a single receiver. The transmitters are further configured with appropriate center frequencies based on their assigned frequency bands. The receiver sensitivity is set to -70 dBm to ensure accurate signal detection, even in challenging environments with potential signal degradation due to multipath effects and noise.

To model realistic scenarios, various sources of noise are incorporated into the simulation. These include inherent transmitter non-linearity, inter-bin interference and thermal noise. Diode response-based envelope correction were also considered. Signal offsets are assigned to receiver samples based on the natural overlap of signals from different transmitters, taking into account their distances. 

\vspace{-0.1in}
\section{Evaluation}
\vspace{-0.1in}
In this section we evaluate \name's performance in simulated and real-world scenarios. 

\subsection{Localization Accuracy}
\begin{figure}[htb]
    \centering
    \vspace{-0.1in}

    \includegraphics[width=0.85\linewidth]{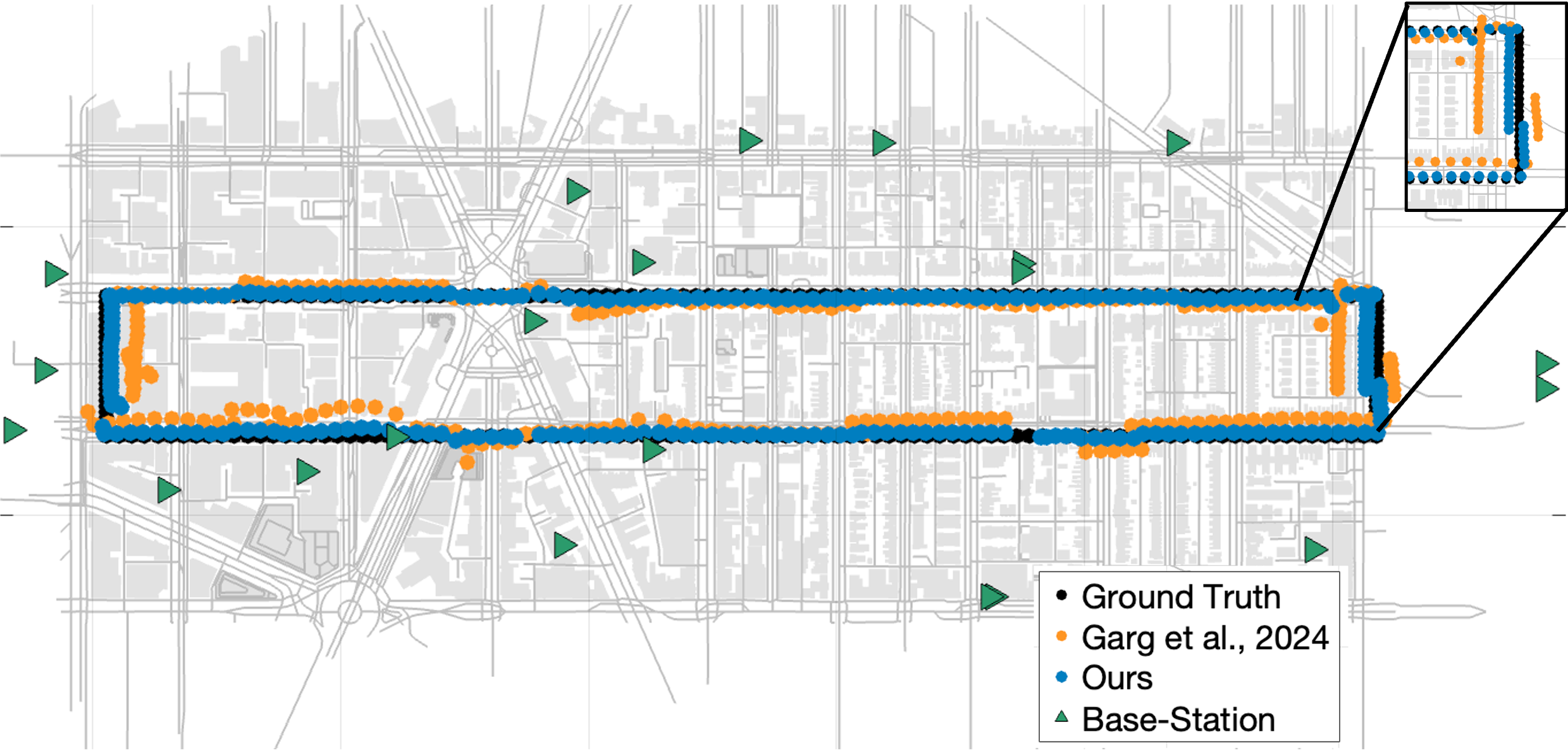}
    
    \vspace{-0.2in}
    \caption{Localization performance of \name vs \cite{garg2024litefoot} in a simulated urban setting.}
    \vspace{-0.1in}
    \label{fig:Simulated_Comparison}
\end{figure}

\vspace{-0.1in}
We compare the localization accuracy of \name against \cite{garg2024litefoot}, which also provides a low-power localization solution using the folded spectrum of LTE signals. In a simulated urban scenario with multiple base stations transmitting LTE signals, we evaluate the performance of both methods along a predefined route. While \cite{garg2024litefoot} achieves a median localization accuracy of 20 meters, \name significantly improves upon this \ref{fig:Simulated_Comparison}, achieving a median accuracy of 4 meters as shown in \ref{fig:Simulated_Comparison_cd}.   .

\begin{table*}[t]
\centering
\caption{Comparison of Positioning Methods in Terms of Localization Error, Energy Consumption, Data Type, and Latency.}
\begin{adjustbox}{max width=\textwidth}
\begin{tabular}{lccccc}
\toprule
\textbf{Method} & \textbf{90\textsuperscript{th} Percentile Error (↓)} & \textbf{50\textsuperscript{th} Percentile Error (↓)} & \textbf{Energy Consumption per inference (↓)} & \textbf{Latency (↓)} & \textbf{Data Type} \\ 
\midrule
\textbf{LTE Compressed Spectrum} &  &  &  &  &  \\ 
\cite{garg2024litefoot} & 45m & 20m & 0.039mJ & 0.01s & Real world \\
\midrule
\textbf{PRS-based System} &  &  &  &  &  \\ 
\cite{dwivedi2021positioning} & 3.4m & - & $~$215mJ \cite{qorvo2024rf2052} & 0.66s & Simulation\\ 
\midrule
\textbf{\name (Ours)}  
 & 2.94m & 2.128m & 3.884mJ & 0.01s & Real world \\ 
\bottomrule
\end{tabular}
\end{adjustbox}
\label{tab:comparison}
\vspace{-0.1in}
\end{table*}

\vspace{-0.1in}
\subsection{Effect of Coarse Estimation}
\begin{figure}[htb]
    \centering
    \vspace{-0.2in}

    \begin{minipage}[t]{0.49\linewidth}
        \centering
        \includegraphics[width=0.9\linewidth]{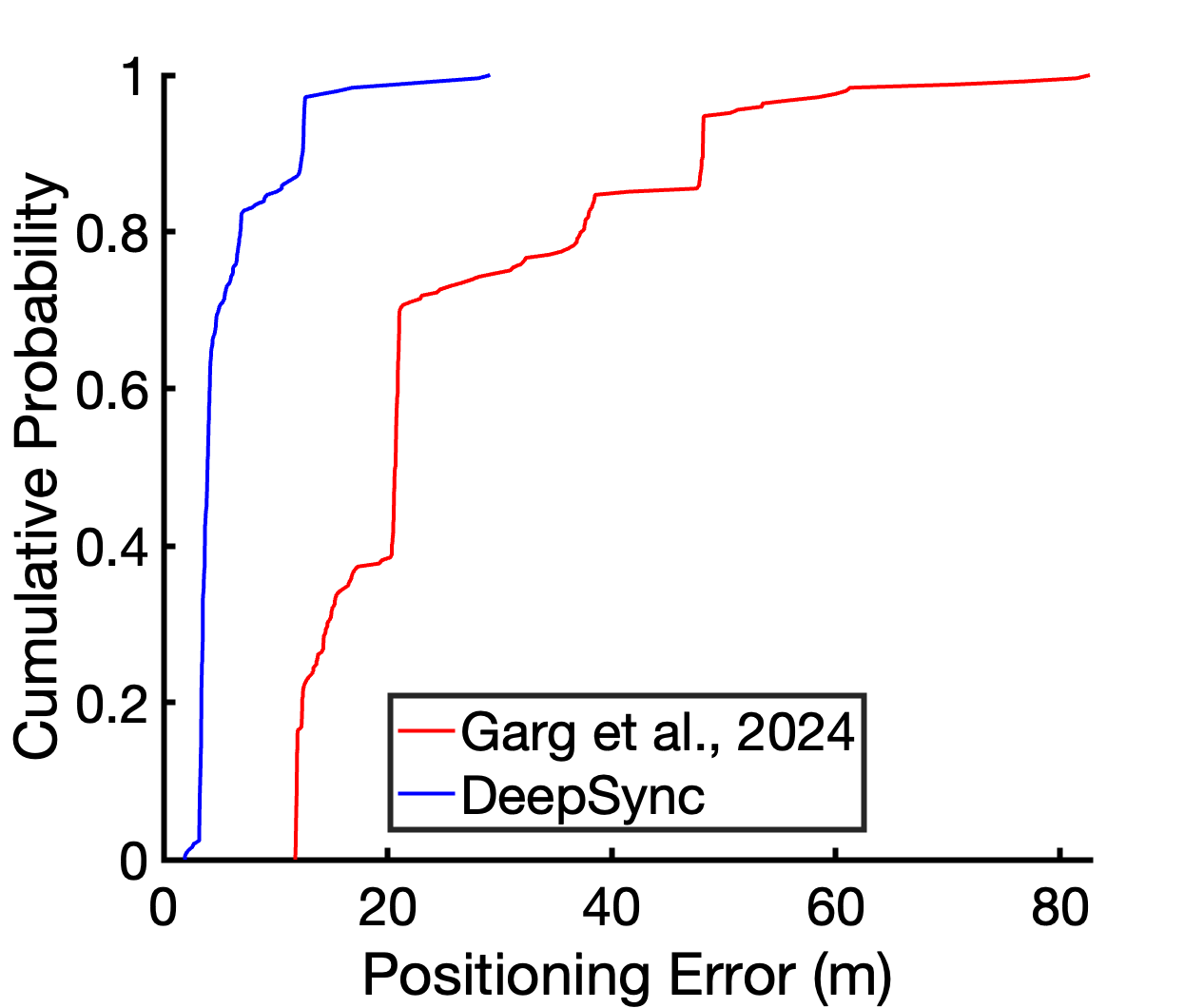}
        \vspace{-0.2in}
        \caption{CDF of localization}
        \label{fig:Simulated_Comparison_cd}
    \end{minipage}
    \hfill
    \begin{minipage}[t]{0.49\linewidth}
        \centering
        \includegraphics[width=\linewidth]{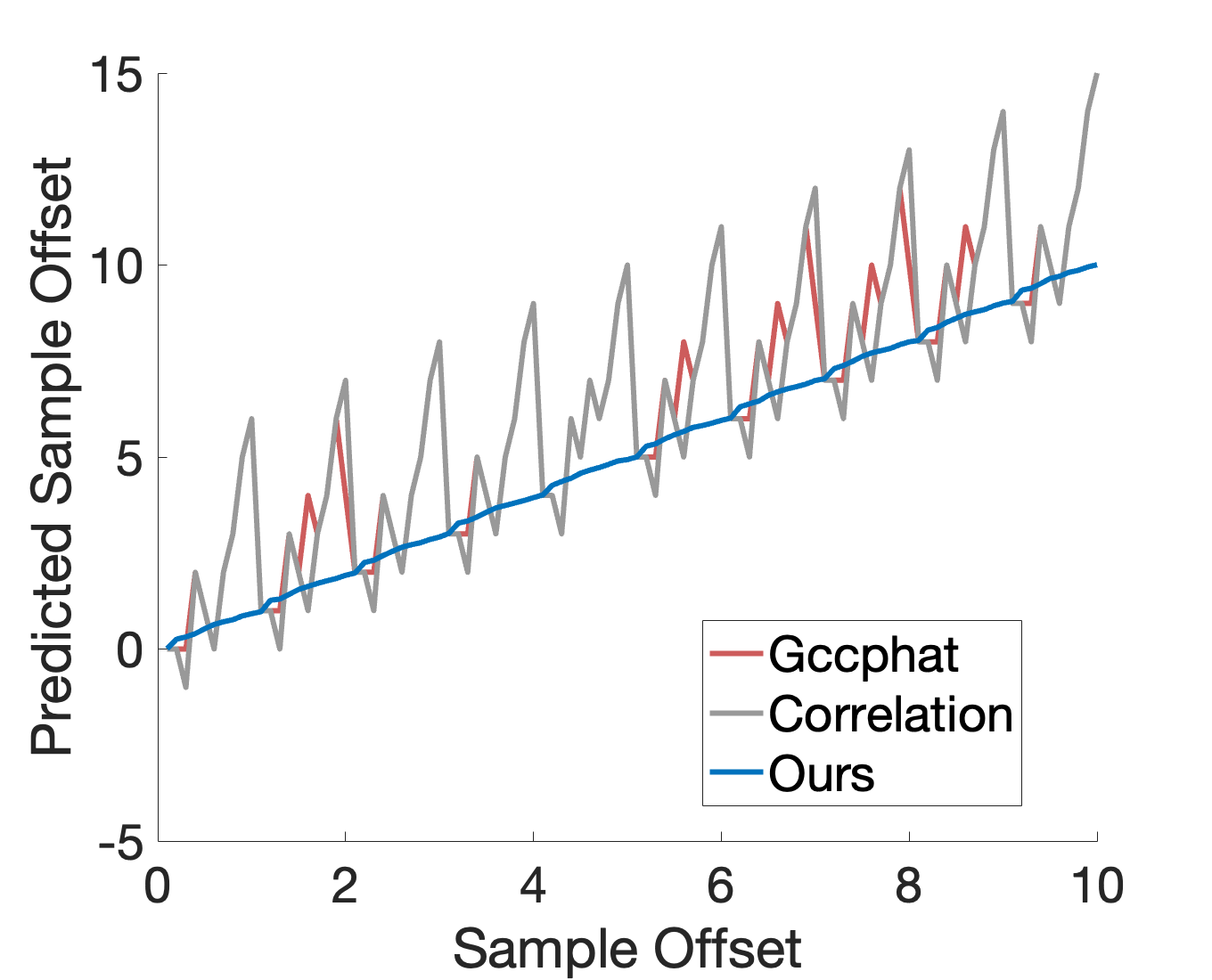}
        \vspace{-0.2in}
        \caption{Offset estimation}
        \label{fig:sample_off}
    \end{minipage}

    \vspace{-0.15in}
\end{figure}

\vspace{-0.1in}
Traditional correlation-based methods in inter-modulated spectrum lack precision, failing to accurately capture sub-sample offsets and introducing spurious errors in sample offset estimation. In contrast, \name effectively estimates sub-sample offsets with a nominal median error of just 0.239. The performance of \name in inter-modulated spectrum is shown in Figure \ref{fig:sample_off}.

\begin{figure}[htb]
    \centering  
    \vspace{-0.1in}

    \parbox{0.49\linewidth}{
        \centering
        \includegraphics[width=\linewidth]{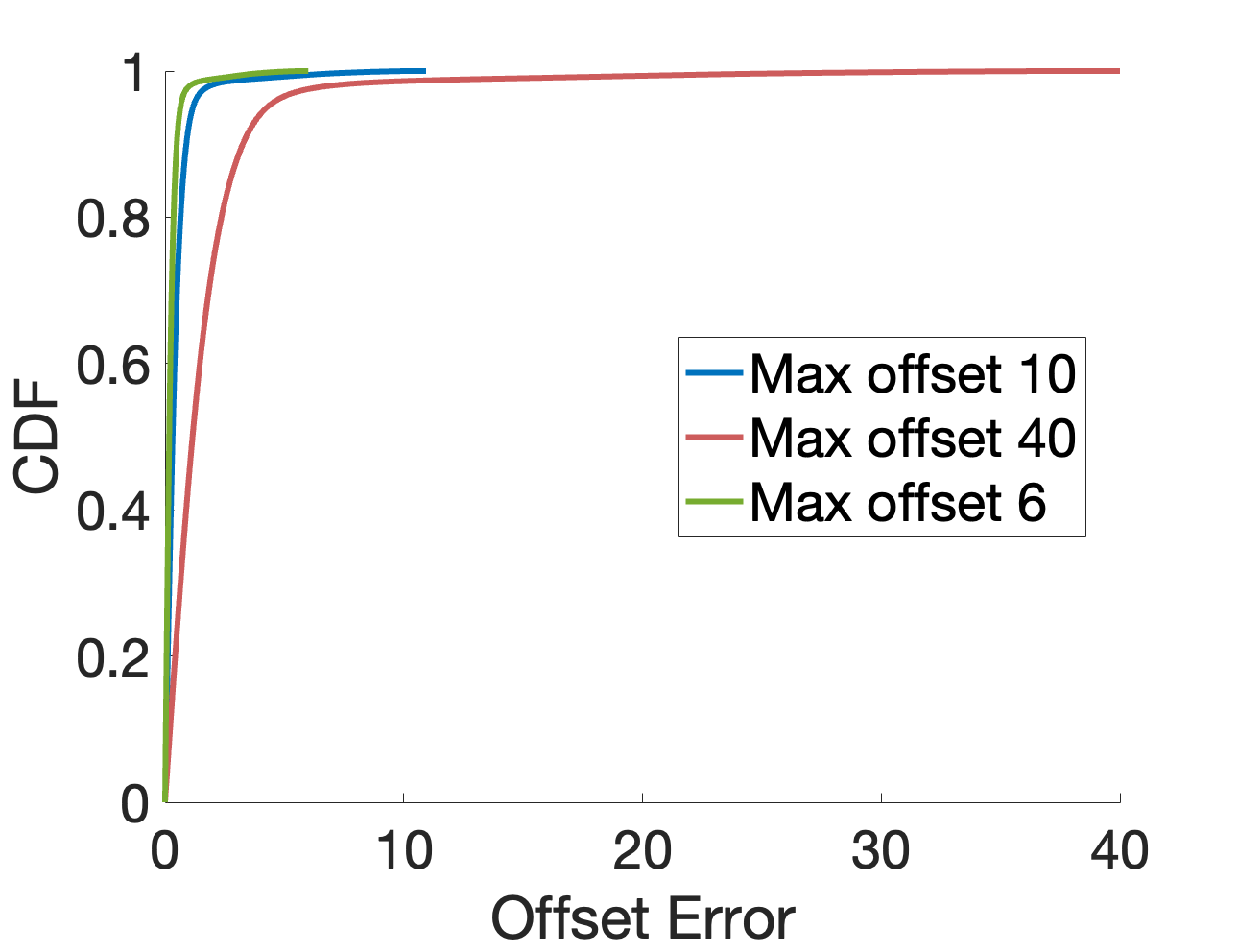}\\
        (a) 
    }
    \hfill
    \parbox{0.49\linewidth}{
        \centering
        \includegraphics[width=\linewidth]{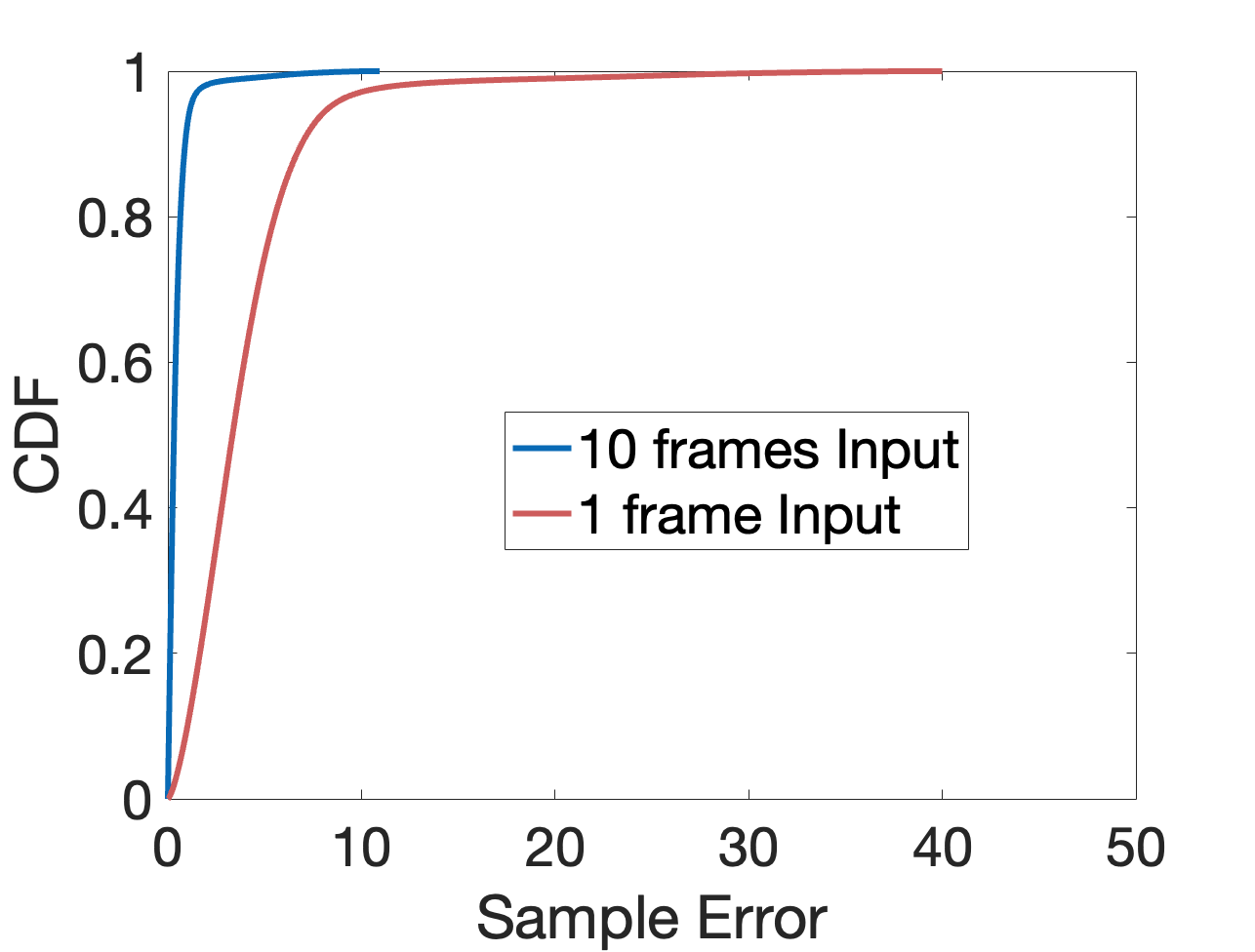}\\
        (b) 
    }

    \vspace{-0.1in}
    \caption{CDF error of sample offset: (a) With different offset size, (b) With different input sizes.}
    \vspace{-0.1in}
    \label{fig:cdf_comparison}
    \vspace{-0.1in}
\end{figure}

Since the choice of the starting index for the signal frame depends on coarse estimation, we evaluated \name’s performance under varying coarse estimation errors, considering maximum erroneous sample offsets of 40, 10, and 6, as shown in Figure \ref{fig:cdf_comparison} (a). The variation in error across these scenarios arises due to the behavior of the Huber loss function, which transitions between quadratic (L2) and linear (L1) penalties based on the error bound. With a larger bound of 40, the model tolerates larger errors by treating them within the L2 region, leading to a higher median error of approximately 0.8 samples. In contrast, a smaller bound of 10 penalizes larger errors more aggressively, encouraging the model to minimize them, thereby reducing the median error to 0.24. Since coarse synchronization is unlikely to introduce an offset error as large as 40, we trained our \name models assuming a random offset error from coarse estimation of 10.

Since \name operates on inter-modulated and spectrum-folded spectrum, it inherently suffers from poor SNR conditions. Training with a single frame results in a median error exceeding 3 samples, which is insufficient for precise localization. However, due to the periodic nature of LTE frames, \name can leverage multiple consecutive frames to improve performance. By incorporating a look-ahead mechanism—where each inference is made using the target frame along with the next 10 consecutive frames—the model significantly reduces the median error by a factor of 15, as shown in Figure \ref{fig:cdf_comparison} (b). This improvement occurs because the model gains a broader temporal view, making it more robust to noise variations and ensuring more accurate offset estimation.

\subsection{Real World Localization}
\vspace{-0.1in}

\begin{figure}[htb]
    \centering
    \vspace{-0.1in}    \includegraphics[height=0.5\linewidth]{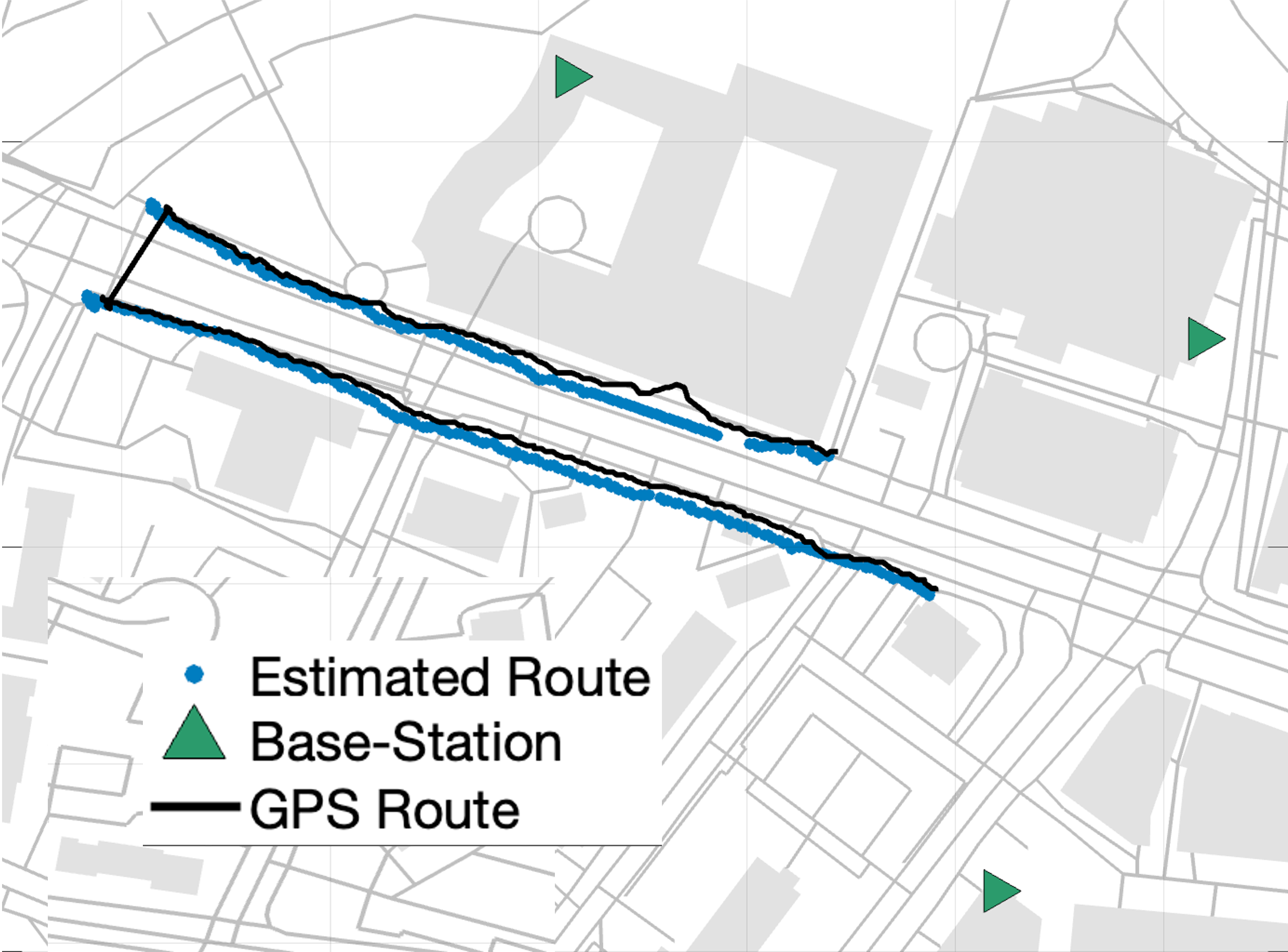} 
    \vspace{-0.1in}
    \caption{Localization performance on real-world data.}
    \vspace{-0.15in}
    \label{fig:Real}
\end{figure}

LTE traces were collected using a USRP N210 \cite{ettus_usrp_n210} along a predefined route, as shown in Figure \ref{fig:Real}. Three Base-stations on the route were operating at 1.932 GHz , 2.115 GHz and 2.145 GHz; after that, the non-linear transformation was theoretically applied to the frames and they were added to simulate the folded spectrum effect.

As the LTE frames captured in three traces were not synchronized, localization traces were manually corrected for the drift. On this real-world dataset, \name achieved a median localization error of 2.128  meters, demonstrating its effectiveness in real world scenarios. The real-world scenario performed better than simulation due to, the simulated route had dense buildings causing multipath interference, and certain locations had low SNR due to sparse tower coverage. Additionally, the simulation used much higher bandwidth to model an envelope detector, while the real-world setup was limited to 45 MHz, impacting performance..

\begin{figure}[htb]
    \centering
    \vspace{-0.1in}

    \parbox{0.49\linewidth}{
        \centering
        \includegraphics[width=\linewidth]{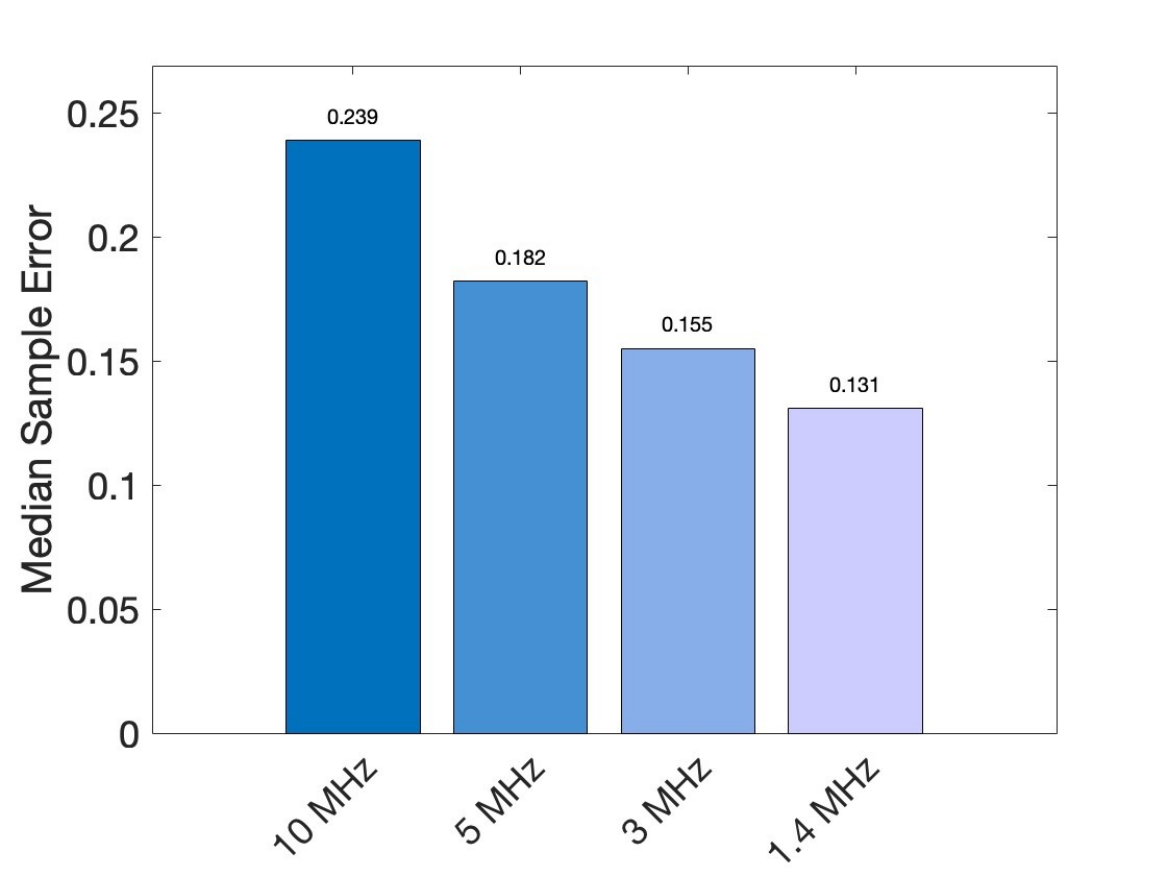}\\
        (a)  .
    }
    \hfill
    \parbox{0.49\linewidth}{
        \centering
        \includegraphics[width=\linewidth]{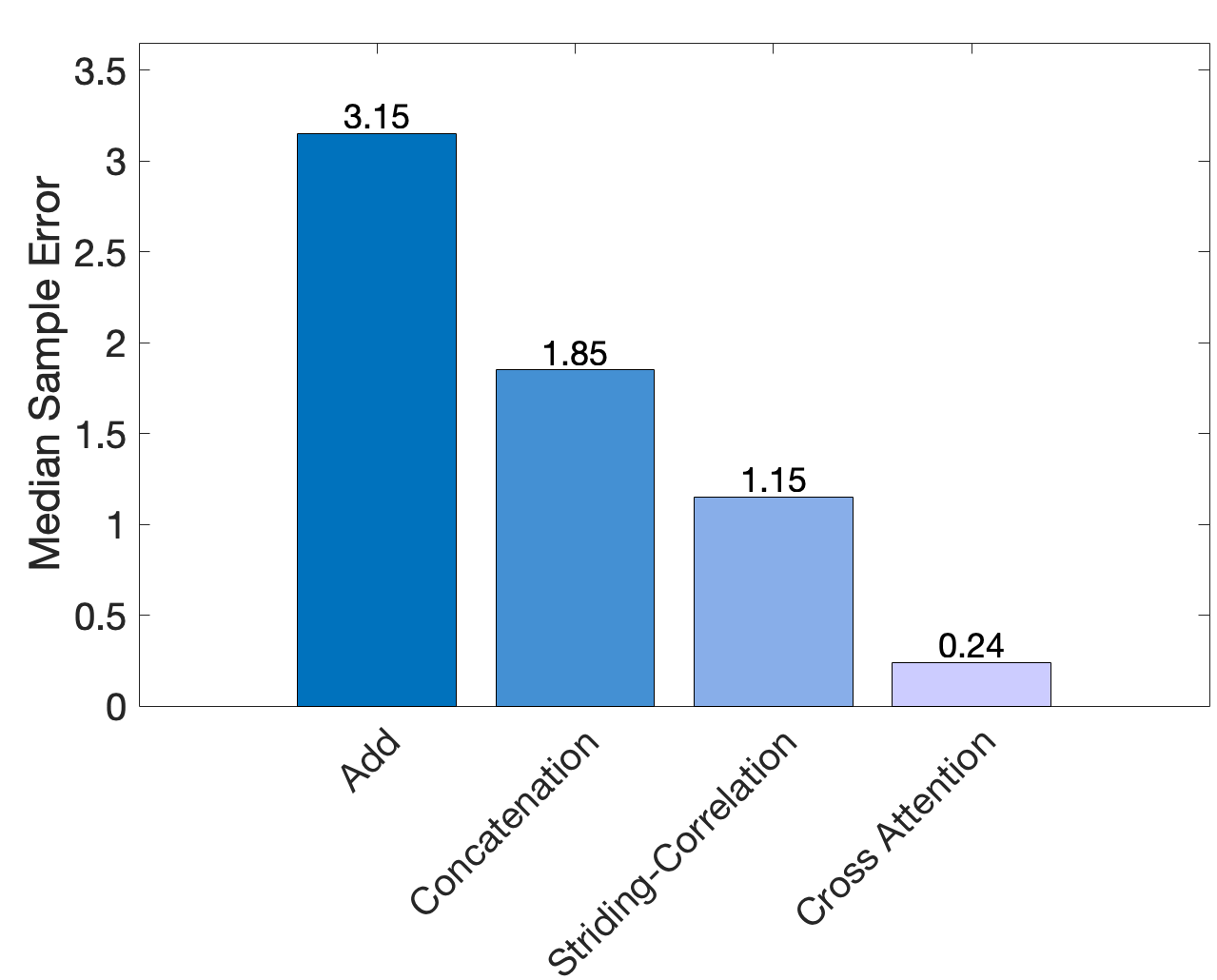}\\
        (b) 
    }

    \vspace{-0.1in}
    \caption{Comparison of different factors affecting \name's performance: (a) Effect of bandwidth on median sample error, (b) Median error for feature combination methods.}
    \vspace{-0.1in}
    \label{fig:median_error}
\end{figure}

The LTE bandwidth directly influences \name’s sub-sample estimation performance. Due to the LTE frame structure, the bandwidth (BW) is inversely proportional to the signal-to-noise ratio (SNR). For instance, a 20 MHz bandwidth corresponds to an SNR of -10 dB, whereas a 5 MHz bandwidth results in an SNR of -5.5 dB, assuming no additional noise sources. As shown in Figure \ref{fig:median_error} (a), higher bandwidth degrades \name’s accuracy in estimating sample offsets due to the increased noise levels. 

\vspace{-0.05in}
\subsection{Feature Combination Methods}
\vspace{-0.1in}

To combine features from the target sync-code and the signal’s latent space after the CNN layers, multiple methods were explored in addition to cross-attention. These included concatenation, element-wise addition, and correlation between feature spaces. It was observed that cross-attention yielded superior results. The median errors for all methods are shown in Figure \ref{fig:median_error} (b). We select cross-attention for its superior performance.

\subsection{Comparison With Other Positioning System}
\vspace{-0.1in}

\name achieves a balance between accuracy, power consumption, and latency. While \cite{garg2024litefoot} demonstrates ultra-low power operation with real-world data, its accuracy remains suboptimal. In contrast, 3GPP standards introduce a specialized signal for localization, the Position Reference Signal (PRS), which enables sub-meter accuracy \cite{r1-1902549}. Methods leveraging PRS, such as those proposed in \cite{dwivedi2021positioning} and \cite{muursepp2021performance}, surpass \name in accuracy but require significantly higher power and latency due to their reliance on traditional, power-intensive RF architectures for decoding PRS signals.

The proposed method strikes a balance between accuracy and energy efficiency, consuming 3.825 mJ of energy while executing 46 million FLOPs per inference on an edge AI device \cite{stm32n6,tu2023unveiling}. Despite this marginal energy overhead, \name achieves a median localization accuracy of 2.128 meters.

\section{Limitation and Future Work}
\vspace{-0.1in}
The results presented in this paper are based on simulated datasets. A real-world implementation of this method would require the system to adapt to several practical challenges. Specifically, it must account for the non-linear transformations introduced by a real-world envelope detector, the transfer functions of low-pass filters and amplifiers, and the frequency-dependent loss inherent in antennas operating across wide bandwidths. Since antennas are frequency-selective devices, their response varies with frequency, introducing additional signal distortion. Furthermore, long-duration experiments would require compensation for clock drift in commercially available ADCs, which can impact synchronization accuracy over time. In future work, we aim to address these challenges to develop a practical, real-world implementation of our system. 

\section{Conclusion}
\vspace{-0.1in}
\name introduces a deep learning-based localization system that leverages sync-code for precise positioning while maintaining ultra-low power consumption. By estimating sub-sample timing offsets through a cross-attention architecture, \name achieves 2.128m median accuracy in real-world tests and 4m in simulations. \name enables precise, power-efficient localization without modifying existing infrastructure, making it a practical solution for asset tracking, geo-fencing, and low-power IoT applications.

\section*{Acknowledgement}
This work was partially supported by NSF CAREER Award 2238433. We also thank the various companies that sponsor the iCoSMoS laboratory at UMD.

\bibliography{main}
\bibliographystyle{icml2025}

\newpage
\appendix
\onecolumn

\end{document}